\documentclass[fleqn,twoside]{article}
\usepackage{espcrc2}

\usepackage{graphicx}
\usepackage[figuresright]{rotating}

\newcommand{\AmS}{{\protect\the\textfont2
  A\kern-.1667em\lower.5ex\hbox{M}\kern-.125emS}}

\title{Theory and Phenomenology of the Polyakov loop in QCD Thermodynamics}

\author{P. Meisinger\address[WU]{Dept. of Physics, 
        Washington University, \\ 
        St. Louis, Mo, 63130, USA}
        T. Miller\address{Indiana University, \\
        Bloomington, In, 47405  USA}
        and
        M. Ogilvie\addressmark[WU]\thanks{Supported by the U.S. Dept. of Energy
		under grant DE-FG02-91ER40628.}
      } 
\begin{document}

\begin{abstract}
A synthesis of Polyakov loop models of the deconfinement transition and
quasiparticle models of gluon plasma thermodynamics leads to a class of
models in which gluons move in a non-trivial Polyakov loop background. These
models are successful candidates for explaining both critical behavior and
the equation of state for the $SU(3)$ gauge theory at finite temperature.
Two sensitive lattice probes which can discriminate between models are the
Polyakov loop $P$ and $\varepsilon -3p$.
\vspace{1pc}
\end{abstract}

\maketitle

%%%% BEGIN MS

The body of results on the critical behavior and thermodynamics
of QCD and related theories at finite temperature are among the most
important that lattice gauge theory simulations have provided, giving
important guidance in both theory and experiment. Theoretical work on the
thermodynamics of the quark-gluon plasma at zero chemical potential is often
benchmarked against results from lattice simulations. 
Here we explore a class of models which synthesize the attractive features
of quasiparticle models of the quark-gluon plasma 
\cite{Peshier:1995ty}\cite{Levai:1997yx}\cite{Schneider:2001nf}
and Polyakov loop models of the deconfinement phase transition 
\cite{Pisarski:2000eq}\cite{Meisinger:2001cq}.
Quasiparticle models, which are partially motivated by perturbation
theory, have been successful in fitting lattice results for
thermodynamic quantities such as the pressure over a
large range of temperatures, but typically at the cost of introducing
a number of extra assumptions and parameters.
Polyakov loop models extend the 
equilibrium free energy density $f$
to a function of the Polyakov loop $P$, which is the 
order parameter for the deconfining phase transition in
pure gauge theories. The spontaneous breaking of
center symmetry in pure gauge theories, 
which underlies our understanding of
the deconfinement transition, occurs in
these models because the free energy
develops symmetry-breaking global minima
above the deconfinement temperature $T_{c}$.

The general class of models of deconfinement we study have
free energies of the form 
\begin{eqnarray}
f(T,P)&=&V(T,P)-p_g(T,P,M(T))\nonumber\\
&&+B(T,P,M(T)).
\end{eqnarray}
The first term $V(T,P)$ is phenomenological, and favors the confined phase
at low temperature. The second term is the negative of the 
quasiparticle pressure $p_g$,
and represents the contribution of gluonic quasiparticles moving in the
presence of a background Polyakov loop. The quasiparticle pressure is given by 
\begin{eqnarray}
p_g(T,P,M(T))&=&-2T\int \frac{d^{3}k}{\left( 2\pi \right)^{3}}\nonumber\\
&&\cdot Tr_{A}\,\ln \left[ 1-e^{-\beta \omega _{k}}P\right] 
\end{eqnarray}
where $Tr_{A}$ denotes the trace in the adjoint representation. The pressure
depends on the quasiparticle mass through the quasiparticle energy, assumed
to have the simple form $\omega _{k}=\sqrt{k^{2}+M\left( T\right) ^{2}};$
the quasiparticle mass is temperature-dependent. The third term is the
so-called bag term. It arises because the zero-quasiparticle state has a
non-zero, temperature-dependent energy $B$. If we assume that $B$ depends on 
$T$ only through the quasiparticle mass $M$, then thermodynamic consistency
leads to an expression for $B\,$which may be written as 
\begin{eqnarray}
B(T,P,M(T))&=&\int_{T_{0}}^{T}d\tau \frac{\partial M^{2} }{%
\partial \tau }\int \frac{d^{3}k}
{\left( 2\pi \right) ^{3}}\frac{1}{k^{2}}\nonumber\\
&&\cdot Tr_{A}\,\ln \left[ 1-e^{-\omega _{k}/\tau }P\right].
\end{eqnarray}
After minimization with respect to $P$, the equilibrium pressure is given by
the negative of the minimum of the free energy density.

In previous work \cite{Meisinger:2001cq}, 
we have developed two models for the $SU(N)$
deconfinement transition which fit within this general class. In both
models, we took $M=0$ and hence $B=0$.
Both of these models gave a good representation of the critical
behavior for all $N$, and reasonable thermodynamic behavior. 
Here we extend the second of
these two models, called model B, to model B+M which includes
quasiparticle mass effects.
The potential \ $V\,$associated with
model B has the form 
\begin{equation}
V(T,P)=-\frac{T}{R^{3}}\ln \left[ \mu \left( P\right) \right] +v_{0} 
\end{equation}
where $\mu (P)$ is Haar measure on the gauge group. This form for
the potential can be understood 
as enforcing color neutrality at scales $R$ and larger. 
We again use the parameter $R$ to set $T_{c}$
to the result of lattice simulations. The constant $v_{0}$ is used to match
the pressure at one point on the curve, again chosen to be $T_{c}$.

\begin{figure}[htb]
\vspace{24pt}
\includegraphics[width=17pc]{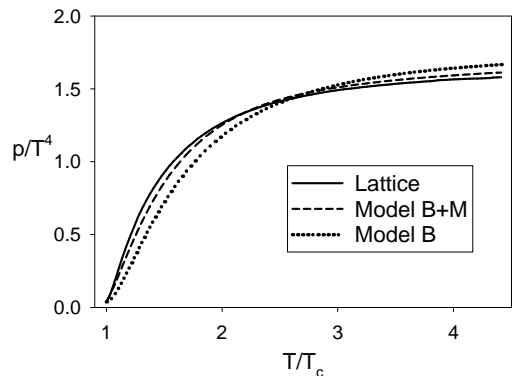}
\vspace{-68pt}
\caption{$p/T^4$ versus $T/T_c$.}
\label{fig:pcross}
\end{figure}

We show in figure \ref{fig:pcross}
the dimensionless pressure \ \ $p/T^{4}\,$as a function of 
$ T/T_{c}$ for models B and B+M, along with simulation data 
from \cite{Boyd:1996bx}.  Note that
once $R$ and $v_0$ are fixed by determining $T_{c}$ and $p(T_c)$, 
there are no more free parameters in model B.
The pressure from Model B overshoots the lattice data at higher
temperatures, and the quasiparticle mass corrects this. 
The quasiparticle mass term has the form suggested by
perturbation theory, $M(T)=g(T)T/\sqrt{2}$.
We take the temperature-dependent
coupling constant to be given by 
$g^2(T)={8\pi ^{2}}/{11\ln \left( T/\Lambda \right) } $
where $\Lambda $ is an adjustable parameter. We have obtained the best fit
to the lattice pressure data by replacing $T/\Lambda $ in $g(T)$ with a
crossover function of the form 
$\left[ \left( T/\Lambda \right) ^{n}+\left( T_{cr}/\Lambda %
\right) ^{n}\right] ^{1/n}$
where we have taken $n=10$. Both Polyakov loop effects and a rapidly rising
quasiparticle mass suppress the temperature as one moves to lower
temperature, and both are not needed. It is physically plausible that $g$
does not run with $T$ all the way down to zero temperature, and the
crossover function provides that effect.
The critical temperature is taken to be $T_{c}=0.272\,GeV$,
although an exact number is not needed. The parameter $R$ is determined from 
$T_{c}$, and is given by $R^{-1}=0.61T_{c}=0.17\,GeV$ which may be
compared with the result for model $B$, $R^{-1}=73T_{c}=0.2\,GeV$. The value
of $\Lambda $ obtained was $0.3T_{c}=.081\,GeV$, with a crossover
temperature $T_{cr}=2.4T_{c}$. 

\vspace{-16pt}

\begin{figure}[htb]
\vspace{24pt}
\includegraphics[width=17pc]{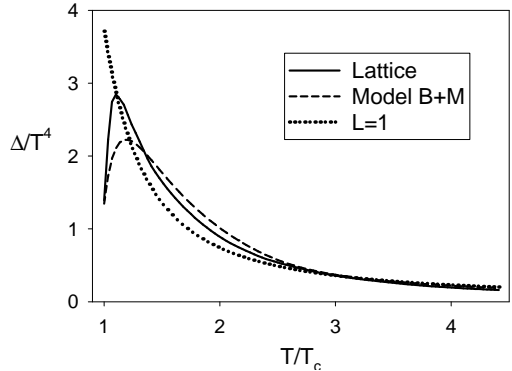}
\vspace{-68pt}
\caption{$(\epsilon - 3 p)/T^4$ versus versus $T/T_c$.}
\label{fig:delta}
\end{figure}

\begin{figure}[htb]
\vspace{8pt}
\includegraphics[width=17pc]{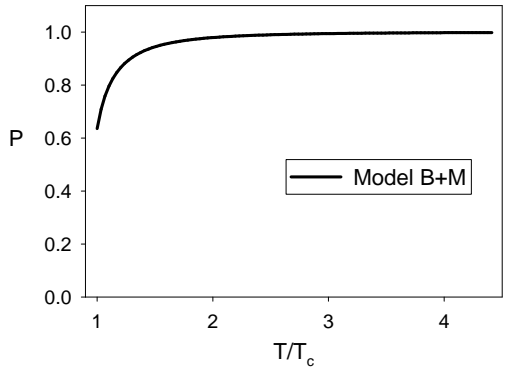}
\vspace{-68pt}
\caption{ $L=Tr_{F}\left( P\right) /N_{c}$ versus $T/T_c$.}
\label{fig:Pol}
\end{figure}

\begin{figure}[htb]
\vspace{24pt}
\includegraphics[width=17pc]{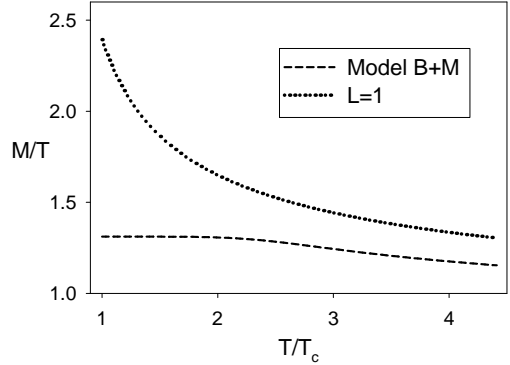}
\vspace{-68pt}
\caption{$M(T)/T$ versus versus $T/T_c$.}
\label{fig:Mred}
\end{figure}

Figure \ref{fig:delta} compares the dimensionless
interaction measure $\Delta =(\varepsilon -3p)/T^{4}$ for model B+M
with both lattice results and a quasiparticle model without Polyakov loop
effects ($L=1$).
Model B+M
shows a peak in $\Delta $ just above $T_{c}$, consistent with the lattice
data, and correctly reproduces the latent heat of the first-order $SU(3)$
deconfinement transition. Of course, the model without Polyakov loops
effects is incapable of reproducing the deconfinement transition. The
first-order character of the $SU(3)$ deconfinement transition is also clear
in 
Figure \ref{fig:Pol}, which shows the Polyakov loop as a function of $T/T_{c}$%
The shape of the curve is similar to that found in a recent
lattice determination of the Polyakov loop \cite{Kaczmarek:2002mc}.
It is also interesting to
compare the quasiparticle masses for model B+M with the $L=1$ model,
as shown in figure \ref{fig:Mred}.
The most notable difference is the rapid rise of the quasiparticle mass
in the $L=1$ model near $T_c$.

%\vspace{-24pt}

\vspace{8pt}

\vspace{-24pt}

It should be apparent that many models will be able to fit lattice
thermodynamic results well. The ability to describe both the thermodynamic
results and the deconfinement transition, while not unique to the model
described here, seems very desirable. The introduction of an extra term $V$
in the free energy is phenomenological, but based on our understanding of
the deconfinement transition. The onus of describing the temperature region
just above $T_{c}$ is placed where it likely belongs, on the confinement
mechanism. It is possible to extend this model to include quarks. 
As with confinement,
our inability to describe chiral symmetry breaking in a fundamental
way forces us to use phenomenological models, in this case 
of Nambu-Jona Lasinio type. Nevertheless, a
useful synthesis of our current understanding of deconfinement and chiral
symmetry restoration is likely possible.

%%%% END MS

\end{document}